\documentclass[12pt]{iopart}

\usepackage{iopams}

\expandafter\let\csname equation*\endcsname\relax

\expandafter\let\csname endequation*\endcsname\relax

\usepackage{amsmath}

\usepackage{graphicx}
\usepackage{float}
\usepackage{cite}
\usepackage{cancel}
\usepackage{hyperref}
\usepackage{subcaption} 
\usepackage{centernot}
\begin{document}

\title[]{Semi-classical understanding of flux quantization in superconductors}

\author{Kolahal Bhattacharya}

\address{St. Xavier's College (Autonomous), Kolkata-700016}
\ead{kolahalbhattacharya@sxccal.edu}
\vspace{10pt}
\begin{indented}
\item[]
\end{indented}

\begin{abstract}
Like electric charge, magnetic flux is also quantised. Theoretically, one can show that the flux quantum must be $h/e$, as observed in the quantum Hall effect. However, in the superconducting systems, the flux quantum is experimentally observed as $h/2e$. In this article, we argue that this phenomenon is fundamentally linked to the nonlocality problem of the Aharonov-Bohm effect and present a new semi-classical explanation for the magnetic flux quantum in superconductivity. This work will also show why the flux quantum should be $h/e$ in the case of the quantum Hall effect.
\end{abstract}
%
%
%
%
%

\section{Introduction}\label{Sec1}
In the first half of the last century, it was experimentally discovered that the magnetic field is expelled from a superconductor during its transition to the superconducting state when it is cooled below the critical temperature. This expulsion pushes off a nearby magnet. This phenomenon is known as the Meissner effect~\cite{meissner1933new}. Later, it was also discovered that the magnetic flux, is quantized within the superconductors in the units of $\frac{h}
{2e}$~\cite{deaver1961experimental} where $h$ is Planck's constant and $e$ is electronic charge. This was shocking since in general the magnetic flux is quantized in the units of $\frac{h}{e}$ (see: Chapter 21 of~\cite{feynman1965feynman},~\cite{doll1961experimental}). The extra factor of $\frac{1}{2}$ was a hint that somehow electrons form a pair in superconductors. This was quite surprising, due to the Coulomb repulsion between the electrons in the pair. At this point, the BCS theory provided a microscopic model of the pairing of electrons, taking into account their phonon-mediated interaction with the superconducting lattice. Several other unconventional superconductors, those for which BCS theory is not applicable, were invented by the end of the last century. Irrespective of whether or not a superconductor is conventional, it exhibits the Meissner effect~\cite{hirsch2012origin}. For unconventional superconductors which are strongly correlated electron systems, presently there is no universally accepted mechanism of electron pairing.

Apart from the introductory chapters in classic popular texts~\cite{feynman1965feynman}, formal treatment of the topic can be found in the standard books~\cite{innes1980introduction,kresin2013fundamentals,tinkham2004introduction}. These texts, as well as review articles~\cite{sharma2015review, alade2022comprehensive} introduce the concept of Cooper pairs of electrons in superconductors. Such pairing can be achieved through the microscopic BCS theory~\cite{cooper1956bound,bardeen1957theory}. In this theory, the total energy reduction by electron pairing has been studied extensively for years and currently it is a calculable quantity. It can be used to predict the superconducting-normal phase transition in many superconductors. But to the best of the knowledge of the author, this empirical fact of electron pairing has never been explained from the perspective of the Meissner effect and Aharonov-Bohm type nonlocality, or similar fundamental grounds. Four decades ago, Crawford~\cite{crawford1982elementary} gave an ``elementary'' derivation of the magnetic flux quantum. But close inspection of his work shows that it is a mere justification of the quantization of magnetic flux, in the units of $\frac{h}{q}$ using the Meissner effect (where $q$ is the electric charge). Unless one already knows about the electron pairs, his reasoning does not explain why the magnetic flux must be quantized in the units of $\frac{h}{2e}$ in superconductors. As such, there is no clear answer to the following question: what makes the magnetic flux equal to $\frac{h}{e}$ in some cases (e.g. quantum Hall effect), and $\frac{h}{2e}$ in the superconductors.



A recent development~\cite{bhattacharya2021demystifying} has presented a local explanation of the nonlocality problem of the Aharonov-Bohm effect in terms of a semi-classical theory of static curl-free fields, somewhat different from the standard QED. In this model, one talks about the wavefunctions of the fields, different from the quantum states of the electromagnetic four potential $A^\mu$. Apart from resolving the nonlocality problem of the Aharonov-Bohm effect, the semi-classical theory also explains the quantization of electric charge~\cite{bhattacharya2023semi} in non-relativistic domain. An interesting prediction of this theory is that wavefunctions of a conservative static field may exist in a region of space as a unitary geometric phase, even if the corresponding classical field vanishes. The non-vanishing nature of the geometric phase under the zero field condition may have been pointed out earlier~\cite{bohm2003geometric, shapere1989geometric}, but the interpretation of the very phase as the wavefunction of the field has perhaps not been done. In the regions devoid of  
classical fields in the Aharonov-Bohm effect, the wavefunctions of the field operate on the electrons as unitary phases.

In this paper, we show that the semi-classical theories of both the magnetostatic field $\vec{\mathcal{H}}$ and the vector potential $\vec{\mathcal A}$ are relevant in superconductors. It is possible to express wavefunctions of these fields which operate on the electrons. From this standpoint, we will try to throw light on flux quantization. We shall start our discussion by summarising the 
basic principles of the semi-classical model of the curl-free vector fields in section~\ref{Sec2}. Relevant examples are vector potential $\vec{\mathcal{A}}$ in a region devoid of the magnetic field, or magnetic field $\vec{ \mathcal{H}}$ in a region devoid of free source current density $\vec{\mathcal J}_{free}$, for which $\nabla\times \vec{\mathcal{H}}={\bf0}$). When an electron, represented by $\psi_e$, is subjected to vector potential $\vec{\mathcal A}$ in the absence of a magnetic field, the wavefunction $\psi_A$ of $\vec{\mathcal{A}}$ field operates on $\psi_e$. We shall show in section~\ref{Sec3} that this electron satisfies the standard Schr$\ddot{\rm o}$dinger equation of a charge under the influence of a vector potential. In the following section~\ref{Sec4}, we will obtain the basis wavefunctions of the magnetic field $\vec{\mathcal H }$ when $\mathcal{H}\neq0$. Then, in section~\ref{Sec5}, we shall first express the basis wavefunction in terms of the vector potential, and then argue why the electrons must form a pair inside superconductors leading to a flux quantum of $\frac{h}{2e}$.

\section{Summary of semi-classical theory of magnetic fields} \label{Sec2}
Under specific conditions, the vector potential $\vec{\mathcal {A}}$, or the magnetic field vector $\vec{\mathcal{H}}$ become curl-free. One can express such a curl-free vector field $\vec {\mathcal{F}}$ as a gradient of a scalar potential $\vec{\mathcal F}=-\nabla{U}$. Such a vector field satisfies a variational principle: $\delta\int\mathcal F ds=0$, where $ds$ denotes a differential length element, similar to the wave vector $\vec{k}$ in optics (see section 53 of~\cite{landau2013classical}). It is understood that the integral is evaluated along a curve that is always superimposed with the direction of $\vec{\mathcal F}$. The similarity of the principle with Fermat's principle of optics suggests the possibility of Hamiltonian formulation of the curl-free fields (similar to the Hamiltonian formulation of geometrical optics~\cite{lakshminarayanan2002lagrangian}) and subsequent quantization. This treatment shows that the momenta conjugate to the coordinates $x,y$ and $z$ are $\mathcal F_x$, $\mathcal {F}_y$ and $\mathcal{F}_z$ respectively. Taken together, $(x, y, z, \mathcal{F}_x, \mathcal{F}_y, \mathcal{F}_z)$ constitute a phase space, every point of which represents a field point characterised by Hamiltonian $H=0$.

Taking this further ahead, one can make a transition to the quantum domain, by requiring that the components of $\vec{\mathcal F}$ are operators. These operate on the wave functions $\psi_F$ of the field through the eigenvalue equation:
\begin{equation}\label{eq1}
\hat{\mathcal F}\psi_F=-i\bar\zeta\nabla\psi_F=\vec{\mathcal{F}}\psi_F
\end{equation}
-where $\bar\zeta=\frac{\zeta}{2\pi}$ is a scale factor which plays the role similar to that of the Planck's constant. This factor has the dimension of $U=-\int\vec{\mathcal F}\cdot d \vec s$. The semi-classical aspects of $\vec{\mathcal{F}}$ is manifested, if $\zeta \centernot\rightarrow{0}$, with respect to $U$ (scalar potential), of the problem. It is understood that the wavefunction $\psi_F$ is a single-valued function of the coordinates~\cite{dirac1981principles,koizumi2022schrodinger}. In the source-free region, $\psi_F$ satisfies:
\begin{equation}\label{eq2}
\bar\zeta^2\nabla^2\psi_F+\mathcal{F}^2\psi_F=0
\end{equation}
The scalar wave equation (Eq.\eqref{eq2}) is the differential equation of the wave function of the field $\vec{\mathcal F}$. The continuum structure of points in the classical phase space does not hold any longer, since $\bar\zeta\centernot\rightarrow 0$ in the quantum domain. Thus, the configuration of a system can be best determined up to the minimum unit $\bar\zeta$ that denotes the area of an elementary cell in the phase space.



A non-normalisable basis solution of Eq.\eqref{eq2} is given by $\psi_F={\bf e}^{i\frac{U}{\bar\zeta}}$. Note that the function $U$ is evaluated by line integral of $\vec{\mathcal{F}}=-\nabla U$ and is independent of the integration path (since $\vec{ \mathcal{F}}$ is conservative, by construction). Thus, $\psi_F$ remains single-valued. This plane-wave solution is valid even if $\mathcal F=0$ and is useful to explain the nonlocality problem in the Aharonov-Bohm effect as noted in~\cite{bhattacharya2021demystifying}. However, in a region where $\mathcal{F}\neq0$, a normalisable solution $\Psi _F$ can be constructed by superposing several such plane waves with appropriate coefficients $u(\mathcal{F})$: $\Psi_F=\int u (\vec{\mathcal F}){\bf e}^{i\frac{U}{\bar\zeta}}d\vec{\mathcal F}$.

If in a region the magnetic field is zero, then $\nabla \times\vec{\mathcal{A}}={\bf0}$. So, a similar treatment would lead to the eigenvalue equation:
\begin{align}\label{eq3}
-i\bar\eta\nabla\psi_A=
\vec{\mathcal{A}}\psi_A
\end{align}
Under the Coulomb gauge condition $\nabla\cdot \vec{\mathcal{A}}=0$, Eq.\eqref{eq3} leads to the following wave equation:
\begin{equation}\label{eq4}
\bar\eta^2\nabla^2\psi_A+\vec{\mathcal{A}}^2\psi_A=0
\end{equation}
It has been shown~\cite{bhattacharya2021demystifying} that the wavefunction $\psi_A$ has the form $\psi_A={\bf e}^{-ie\frac {\int{\mathcal A} d{s}}{\hbar}}$ where $\bar\eta =-\frac{\hbar}{e}$. However, in the present context, it is more meaningful to use a closed loop line integral in the exponent: $ \psi_A=Exp\left[{\frac{i}{\bar \eta}\oint\vec{\mathcal{A}}\cdot d{\vec s}}\right]$. To show this, we start from Eq.\eqref{eq3}:
\begin{align}\label{eq5}
&-i\bar\eta\nabla\psi_A=\vec
{\mathcal{A}}\psi_A\nonumber\\
\implies& -i\bar\eta\nabla(\ln\psi_A)=\vec
{\mathcal{A}}\nonumber\\
\implies& -i\bar\eta\ \nabla(\ln\psi_A)\cdot d{\vec s}=-i\bar\eta\ d(\ln\psi _A)=\vec
{\mathcal{A}}\cdot d{\vec s} \nonumber\\
\implies& -i\bar\eta\oint d(\ln \psi_A)=\oint\vec{\mathcal{A}}\cdot d{\vec s}\nonumber\\
\implies&\ln\psi_A=\frac{i}{\bar\eta}\oint\vec{\mathcal{A}}\cdot d{\vec s}\nonumber\\
\implies&\psi_A=Exp\left[{\frac{i}{\bar\eta}\oint\vec{\mathcal{A}}\cdot d{\vec s}}\right]
\end{align}
The expression without a $\oint$ corresponds to a non-specific lower limit of the integral in the left-hand side. A general normalisable solution of Eq.\eqref{eq5} is constructed as: $\Psi_A=\int a(\vec{\mathcal{A}}){\bf e}^{-i\frac{e\oint\vec {\mathcal{A}}\cdot d\vec{s}}{\hbar}} d{\vec{\mathcal{A}}}$. 

\section{Transformation of electron state in zero magnetic field}\label{Sec3}
First, let us assume that in the absence of a magnetic field, only the wavefunction $\Psi_A$ representing the vector potential $\vec{\mathcal
A}$ operates on a given electron. The electron wave packet $\psi_e$ is transformed as:
\begin{equation}\label{eq6}
\psi_e\rightarrow\psi'_e=\Psi_A\psi_e=\left[\int a(\vec{\mathcal A}){\bf e}^{-i\frac {e\oint\vec{\mathcal A}\cdot d{\vec{s}} }{\hbar}}d{\vec{\mathcal A}}\right]\psi_e
\end{equation}
Time variation of the transformed state is given by:
\begin{align}\label{eq7}
i\hbar\frac{\partial\psi_e'}{\partial t}=i\hbar\left[\int a(\vec{\mathcal{A}}){\bf e}^{-i\frac{e\oint\vec{\mathcal{A}}\cdot d{\vec s}} {\hbar}} \frac{\partial}{\partial t}\left(-i\frac{e\oint{\vec{\mathcal{A}}}\cdot d{\vec s}}{\hbar}\right) d{\vec{\mathcal{A}}}\right]\psi_e+i\hbar \Psi_A\frac{\partial\psi_e}{\partial t}=\Psi_A\left(i\hbar \frac{\partial\psi_e}{\partial t}\right)
\end{align}
Here we can drop the first term, taking into account the fact that the partial temporal derivative of $\oint\vec{\mathcal{A}}\cdot d\vec{s}$ is zero, as far as magnetostatics is concerned. On the other hand, by expressing the normalized electron wavefunction as $\psi_e=\int c(\vec{k}){\bf e}^{i({\vec{k}}\cdot{\vec{r}}-\omega t)} d{\vec{k}}$, the kinetic term of the transformed state is given by:
\begin{align}\label{eq8}
    -\frac{\hbar^2}{2m}\nabla^2\psi_q'&=-\frac{\hbar^2}{2m}\left(\nabla^2\Psi_A\psi_q+2\nabla\psi_A\cdot\nabla \psi_q\right)+\Psi_A\left(-\frac{\hbar^2}{2m}\nabla^2\psi_q\right)\nonumber\\
\implies\Psi_A\left(-\frac{\hbar^2}{2m}\nabla^2\psi_e\right) &= -\frac{\hbar^2}{2m}\nabla^2\psi_e'+\frac{\hbar^2}{2m}\left(-\frac {e^2\vec{\mathcal{A}}^2}{\hbar^2}\Psi_A\right)\psi_e+\frac{\hbar^2}{2m}2(-ie)\frac{\vec{\mathcal{A}}\Psi_A}{\hbar}\cdot(i\vec{k})\psi_e\nonumber\\
&=\frac{1}{2m}\left[-\hbar^2\nabla^2\psi_e'-e^2\vec{\mathcal{A}}^2\psi_e'+2e\hbar(\vec{\mathcal{A}}\cdot\vec{k})\psi_e'\right]
\end{align}
Let us note that:
\begin{align}\label{eq9}
    -i\hbar\nabla(\Psi_A\psi_e)&=-i\hbar\nabla\Psi_A\psi_e-i\hbar \Psi_A\nabla\psi_e\nonumber\\
    &=-i\hbar\frac{-ie\vec{\mathcal{A}}}{\hbar}(\Psi_A\psi_e)-i\hbar(i\vec{k})(\Psi_A\psi_e)\nonumber\\
\implies 2e\vec{\mathcal{A}}\cdot\hat{\vec{p}}\psi_e'&=-2e^2\vec{\mathcal{A}}^2\psi_e'+ 2e\hbar\vec{\mathcal{A}}\cdot\vec{k}\psi_e'
\end{align}
From Eq.\eqref{eq8} and Eq.\eqref{eq9}, we find that:
\begin{align}\label{eq10}
    \Psi_A\left(-\frac{\hbar^2}{2m}\nabla^2\psi_e\right)&=\frac{1}{2m}\left[-\hbar^2\nabla^2\psi_e'+e^2\vec{\mathcal{A}}^2\psi_e'+2e\vec{\mathcal{A}}\cdot\hat{\vec{p}}\psi_e'\right]\nonumber\\
    &=\frac{1}{2m}\left[-\hbar^2\nabla^2\psi_e'-ie\hbar\cancelto{0}{(\nabla\cdot\vec{\mathcal{A}})}\psi_e'-2i\hbar e\vec{\mathcal{A}}\cdot\nabla\psi_e'+e^2\vec{\mathcal{A}}^2\psi_e'\right]\nonumber\\
    &=\frac{1}{2m}\left[(\hat{\vec{p}}+e\vec{\mathcal{A}})^2\right]\psi_e'
\end{align}
Hence, from Eq.\eqref{eq7} and Eq.\eqref{eq10}, we find that the transformed state follows the following equation:
\begin{equation}\label{eq11}
    i\hbar\frac{\partial\psi_e'}{\partial t}=\frac{1}{2m}(\hat{\vec{p}}+e\vec{\mathcal{A}})^2\psi_e'
\end{equation}
This exercise shows that the semi-classical theory of magnetostatic fields is consistent with standard formalism.
\section{wavefunctions of magnetic field}\label{Sec4}
If in a region, the externally applied magnetic field $\vec{\mathcal{H}}$ is non-zero, but the source current $\vec{\mathcal{J}}_{free}$ is zero, we have $\nabla \times\vec{\mathcal H}={\bf0}$. So, one has $\delta\int \mathcal{H}\ ds=0$ when the integral is evaluated along a curve superimposed with the local direction of the magnetic field. Hence, one can develop a semi-classical description of the field $\vec{\mathcal{H}}$. The wavefunctions $\psi_H$ representing the field are seen to satisfy equations similar to Eq.\eqref{eq1} and Eq.\eqref{eq2}:
\begin{align}\label{eq12}
-i\bar\kappa\nabla\psi_H=\vec{\mathcal{H}}\psi_H\hspace{4.0cm} \bar\kappa^2\nabla^2\psi_H+\vec{\mathcal{H}}^2\psi_H=0
\end{align}
Clearly, the basis wavefunction $\psi_H$ has the form $\psi_H={\bf e}^{i\frac{\oint\vec{\mathcal H}\cdot d\vec{s}}{\bar\kappa}}$. The normalizable wavefunctions are $\Psi_H=\int c(\vec{\mathcal{H}}) {\bf e}^{i\frac{\oint\vec{\mathcal H}\cdot d \vec{s}}{\bar\kappa}}d{\vec{\mathcal{H}}}$.

Now, for a superconductor subjected to an external magnetic field $\vec{\mathcal{H}} $, $\nabla\times\vec{\mathcal{H}}={\bf0}$ inside the bulk of the material, since the source current density is located outside. Not only that, there is no bound current $ \vec{\mathcal{J}}_b$ either. There is only a current distribution develops at its surface which opposes the field $\vec {\mathcal{H}}$. Since the bound current density vanishes, the curl of the magnetization vector $\vec{\mathcal{M}}$ is zero as well: $\nabla\times\vec{\mathcal{M}}={\bf0}$. As the total magnetic field is a sum of these two vector fields $\vec{\mathcal{B}}=\mu_0 (\vec{\mathcal{H}}+\vec{\mathcal{M}})$, it follows that the curl of magnetic field is zero as well $\nabla\times\vec{\mathcal{B} }={\bf0}$. Due to the Meissner effect, we know that $\mathcal{B}=0$, but we do not need to use that information at this stage. All we say, is that the magnetic field $\vec{ \mathcal{B}}$ should have a semi-classical model inside the superconductors.
\begin{align}\label{eq13}
-i\bar\beta\nabla\psi_B=\vec{\mathcal{B}}\psi_B\hspace{4.0cm} \bar\beta^2\nabla^2\psi_B+\vec{\mathcal{B}}^2\psi_B=0
\end{align}
-where $\beta$ is a parameter whose value must be determined and $\psi_B$ is a wavefunction that represents the magnetic field $\vec{\mathcal{B}}$. Like before, we can show that the basis wavefunctions are $\psi_B={\bf e}^{i\frac{\oint\vec{\mathcal B}\cdot d \vec{s}}{\bar\beta}}$ and the normalizable wavefunctions are given by $ \Psi_B=\int b(\vec{\mathcal{B}}){\bf e}^{i \frac{\oint\vec{\mathcal B}\cdot d \vec{s}}{\bar\beta}}d{\vec{\mathcal{B}}}$. In the following, we show that these wavefunction can be expressed in an alternative form which is more useful for our future discussion.

The linear momentum of a particle of mass $m$ and charge $q$ moving at a constant speed $v$ in a region of non-zero magnetic field is: $\vec{p}=m\vec{v}+q\vec{\mathcal{A}}$. This implies:
\begin{align}\label{eq14}
\nabla\times\vec{p}&=m\nabla\times\vec{v}+q\nabla\times\vec{\mathcal{A}}\nonumber\\
    {\bf 0}&=2m\vec{\omega}+q\vec{\mathcal{B}}\nonumber\\
\implies\vec{\mathcal{B}}&=-\frac{2m}{q}\vec{\omega}
\end{align}
where $\vec{\omega}$ is the angular velocity. Using this expression, we shall evaluate the integral $\oint\mathcal Bds$ along a curve $\mathcal C$ aligned with the local direction of $\vec{\mathcal B}$, say the $z$ direction. We assume that the motion of the charge produces a constant magnetic dipole moment: $\vec{\mu}$. If the velocity $\vec v$ of the particle locally makes an angle $\theta$ with the $z$ direction, then $ds=v\cos\theta\ dt$. So, the integral can be expressed as:
\begin{align}
    \oint\mathcal{B}ds&=-\oint\frac{2m}{q}\omega\ ds\nonumber\\
                     &=-\oint\frac{||\vec{\mathcal{L}}||}{||\vec{\mu}||}\frac{d\phi}{dt} v\cos\theta\ dt\label{eq15}\\
                     &=-\frac{v\cos\theta}{||\vec{\mu}||}\oint{||\vec{\mathcal{L}}||\ d\phi}\label{eq16}
\end{align}
-where in Eq.\eqref{eq15}, we have used the well-known relation between angular momentum vector $\vec{\mathcal{L}}$ and the magnetic dipole moment of the test current distribution $\vec\mu$, i.e. $\vec{\mathcal L}=\frac{2m}{q}\vec{\mu}$. 

Now, the minimum value of $\bar\beta\equiv\oint\mathcal Bds$ occurs, if we set the minimum allowed value of the action $\oint||\vec{\mathcal L} ||\ d\phi$ to be $\hbar$ and $\cos\theta=1$. Then, from Eq.\eqref{eq16}, we see that  $\bar\beta=-\frac{v\hbar} {||\vec{\mu}||}$. Therefore, the wavefunction $\psi_B$ can be expressed as:

\begin{align}\label{eq17}
    \psi_B=Exp\left[-i\frac{||\vec {\mu}||\oint\mathcal{B}\ v\cos\theta\ dt}{v\hbar}\right]
    =Exp\left[-i\frac{\oint\vec{\mu}\cdot\vec{\mathcal B}dt}{\hbar} \right]
\end{align}
We note that even when $\vec{\mathcal{B}}={\bf0}$, we can still conceive the basis wavefunction $\psi_B={\bf e}^{i\frac{\oint\vec{\mathcal B}\cdot d \vec{s}}{\bar\beta}}$, as observed in~\cite{bhattacharya2021demystifying}. One can identify this as the unitary time evolution operator of a magnetic dipole moment inside a magnetic field.

\section{Flux quantization in superconductors}\label{Sec5}
Now, let us investigate how the electrons form pairs in the superconductors. To understand this aspect, we consider a chunk of superconducting material to which an external uniform magnetic field $\vec{\mathcal{H}}$ is applied, say in the $z$ direction. Now, the temperature is gradually reduced below the critical temperature such that the magnetic field is expelled from the chunk and we have a superconductor. The electrons in this chunk are in the field of $\vec{\mathcal A}$. Since $\nabla\times\vec{\mathcal A} =0$, the electrons are subjected to $\Psi_A=\int a(\mathcal{A}) {\bf e}^{-i\frac{e\oint{\mathcal{A}}d{s}}{\hbar}} d{\mathcal A}$. 
On the other hand, $\vec{\mathcal{B}}={\bf0}$ in this chunk due to the Meissner effect. Though the electrons in this bulk superconducting region are not subjected to macroscopic magnetic field, these could experience the semi-classical wavefunction $\psi_B$ of the same as found in the last section. 

If we assume that the dipoles, created due to the motion of the electrons, are aligned with the direction of the magnetic field ($z$ direction), then $\vec{\mu}\cdot\vec{\mathcal B}=\mu \mathcal{B}$. Noting that $\vec{\mu}=\frac{1}{2}\int _\tau{\bf r}\times\vec {\mathcal J}d\tau$, $\vec{\mathcal J}= \rho\vec{v}$ and the charge by $e=\rho d\tau$, we see that $\vec{\mu}=\mu\hat{z}=\frac{e}{2}{\bf r}\times\vec{v}=\frac {1}{2}e(xv_y-yv_x)\hat{z}$. Hence, $ \vec{\mu}\cdot\vec{\mathcal{B}}=\mu\mathcal{B}=\frac{1}{2}e( xv_y-yv_x)B$. Now, for uniform magnetic field $\vec{\mathcal {A}}= \frac{1}{2}\vec{\mathcal{B}}\times{\bf r}$. If we remove the vector notation, $\mathcal{A}=\frac{1}{2}\mathcal{B}r$, then $\psi_B$ can be written as:
\begin{align}\label{eq18}
\psi_B&=Exp\left[-i\frac{\oint\frac{e}{2}(xv_y-yv_x)\frac{2\mathcal{A}}{r} dt}{\hbar}\right]\nonumber\\
&=Exp\left[-i\frac{\oint e\mathcal{A}\left(\frac{x\cdot dy-y\cdot dx}{\sqrt{x^2+y^2}} \right)} {\hbar}\right]\nonumber\\
&=Exp\left[-i\frac{\oint e\mathcal{A}r d\phi}{\hbar}\right]\nonumber\\
&=Exp\left[-i\frac{\oint e\mathcal{A}ds}{\hbar}\right]
\end{align}
Hence, there are two different wavefunctions acting on the charged entities in a superconductor. One part was discussed in section~\ref{Sec3} and the other one is $\psi_B$ just noted before. Therefore, the effective wavefunction operating on the electrons in superconductors in the Meissner state is given by $\Psi_{AB}= \Psi_A\psi_B= \int a(\mathcal{A}){\bf e}^{-2i\frac{e\oint {\mathcal {A}}d{{s}}} {\hbar}}d{\mathcal{A}}$. This may raise a concern that the construction of an effective wavefunction $\Psi_A\psi_B$ can lead to an arithmetical coincidence of having $-2e$. The concern is apparently justified, as the form of the basis vector $\psi_B$ found in Eq.\eqref{eq18} is identical to the basis vector for $\Psi_A$, as found in Eq.\eqref{eq5}.

This point is addressed in the following. If we limit ourselves to $\vec{\mathcal B}={\bf0}$ case, the equations for $\Psi_A$ and $\psi_B$ are given by:
\begin{subequations}
\begin{align}
\bar\eta^2\nabla^2\Psi_A+\vec{\mathcal{A}}^2\Psi_A&=0\label{eq19a} \\
\nabla^2\psi_B=0\label{eq19b},
\end{align}
\end{subequations}
Putting $\mathcal B=0$ in Eq.\eqref{eq13}. The vector spaces of the two wavefunctions $\Psi_A$ and $\psi_B$ are exclusive, since there exists a gauge freedom in $\vec{\mathcal{A}}$, for all values of $\vec {\mathcal{B}}$. If there were a constraint relation connecting the two vector fields (without any gauge freedom), then perhaps such a factorization could not be achieved. In addition to that, the semi-classical model for the vector potential and the resulting $\Psi_A$ can be found only when $\vec{\mathcal B}={\bf0}$. On the other hand, a normalizable wavefunction for $\Psi_B$ can be conceived for a non-zero magnetic field $\vec{\mathcal{B}}$ if the source current density vanishes, but then $\Psi _A$ cannot be constructed since $\nabla\times \vec{\mathcal{A}}\neq{\bf0}$.

Suppose, we denote the states by the Dirac notation:
\begin{align}\label{eq20}
|\Psi_A\rangle=\sum_i a_i|i\rangle
\hspace{4.0cm}
|\psi_B\rangle=|j\rangle
\end{align}
Though the representative of the basis ket $|j\rangle$ happens to assume the same form as that of $|i\rangle$ in a particular case (the sameness of Eq.(\eqref{eq5}) and Eq.\eqref{eq18}), it must be realised that these basis states belong to different vector spaces. We also see that there is no summation over $j$ in the expression for $|\psi_B\rangle$, since there are no magnetic field values to sum over. Therefore the state $\Psi_{AB}$ can always be factorized:
\begin{align}\label{eq21}
\Psi_{AB}=\sum_i a_i(|i\rangle\otimes|j
\rangle)=(\sum_i a_i|i\rangle)\otimes|j \rangle=|\Psi_A\rangle\otimes|\psi_B\rangle
\end{align}
This shows that these two states are not entangled and that the state $\Psi_{AB}$ can be expressed as $\Psi_A\psi_B$.

Now, we impose the condition that the vector potential remains unchanged when we add a constant $\vec{\mathcal{A}_0}$ to it: i.e. $\vec{\mathcal A}\rightarrow{\vec {\mathcal{A}}+\vec{\mathcal{A}_0}}$.
Then we have:
\begin{align}
a(\vec{\mathcal{A}}+\vec{\mathcal{A}_0})&=a(\vec{\mathcal{A}})\label{eq22}\\
    {\bf e}^{-i\frac{2e\oint(\vec{\mathcal{A}}+\vec{\mathcal{A}_0})\cdot d{\vec{s}}}{\hbar}}&={\bf e}^{-i\frac{2e\oint\vec{\mathcal{A}}\cdot d{\vec{s}}}{\hbar}} \label{eq23}
\end{align}
In fact, for the system we considered, the vector potential can be shown to be $\propto\frac{1}{r}$ in plane polar coordinates. Hence, the transformation by a constant vector potential $\vec{ \mathcal{A}}_0$ effectively implies a transformation within the same circle (of the same radius). Thus, Eq.\eqref{eq22} means that for any such transformation $\vec{\mathcal{A}}\rightarrow{\vec{\mathcal{A}}+ \vec{\mathcal{A}_0}}$, the coefficient $a(\vec{\mathcal{A}})$ remains the same. This reasoning is applicable also in the case of the magnetic Aharonov-Bohm effect, where two $\Psi_A$-s from disjoint paths of equal radius interfere~\cite{bhattacharya2021demystifying}.

On the other hand, Eq.\eqref{eq23} leads to a flux quantization condition, as shown below. Collecting the argument of the cosine part of both sides of the Eq.\eqref{eq23}, we find that:
\begin{align}\label{eq24}
\left(\frac{2e\oint(\vec{\mathcal{A}}+\vec{\mathcal{A}_0})\cdot d{\vec{s}}}{\hbar}\right)&=\left(\frac{2e\oint\vec{\mathcal{A}}\cdot d{\vec{s}}}{\hbar}\right)+2k\pi\ ... ... ... ...\rm{where\ k\in\mathbb{N}}\nonumber\\
\implies 2e\oint\vec{\mathcal{A}_0}\cdot d\vec{s}&=k\ 2\pi\frac{h}{2\pi}=k\ h\nonumber\\
\implies\oint\vec{\mathcal{A}_0}\cdot d\vec{s}&=k\frac{h}{2e}
\end{align}
The above exercise demonstrates that in superconductors under the Meissner state, the magnetic flux is quantized in the integral multiples of $\frac{h}{2e}$. This in turn, implies that charge must be quantized in the units of $2e$. It can be achieved through the model proposed in the BCS theory. 
\section{Validation of the model}\label{Sec6}
We note that the description that the effective wavefunction acting on an electron state is $\Psi_{AB}$ is different from the one discussed earlier in section~\ref{Sec3}. We assert that the wavefunction representing a charged entity $q$ in a type-I superconductor gets transformed as:
\begin{equation}\label{eq25}
\psi_q\rightarrow\psi'_q=\Psi_{AB}\psi_q=\Psi_A\psi_B\psi_q=\left[\int a(\vec{\mathcal A}){\bf e}^{-i\frac {2e\oint\vec{\mathcal A}\cdot d{\vec{s}} }{\hbar}}d{\vec{\mathcal A}}\right]\psi_q
\end{equation}
Time variation of the transformed state is given by:
\begin{align}\label{eq26}
    i\hbar\frac{\partial\psi_q'}{\partial t}&=i\hbar\left[\int a(\vec{\mathcal{A}}){\bf e}^{-i\frac{2e\oint\vec{\mathcal{A}}\cdot d{\vec s}} {\hbar}} \frac{\partial}{\partial t}\left(-i\frac{2e\oint{\vec{\mathcal{A}}}\cdot d{\vec s}}{\hbar}\right) d{\vec{\mathcal{A}}}\right]\psi_q+i\hbar \Psi_{AB}\frac{\partial\psi_q}{\partial t}\nonumber\\
    &=\Psi_{AB}\left(i\hbar \frac{\partial\psi_q}{\partial t}\right)
\end{align}
Like section~\ref{Sec3}, we can drop the first term, taking into account the fact that the partial temporal derivative of $\oint\vec{\mathcal{A}}\cdot d\vec{s}$ is zero, as far as magnetostatics is concerned. Similarly, the kinetic term of the transformed state is given by:
\begin{align}\label{eq27}
    -\frac{\hbar^2}{2m}\nabla^2\psi_q'&=-\frac{\hbar^2}{2m}\left(\nabla^2\Psi_{AB}\psi_q+2\nabla\Psi_{AB}\cdot\nabla \psi_q\right)+\Psi_{AB}\left(-\frac{\hbar^2}{2m}\nabla^2\psi_q\right)
\end{align}\label{eq28}
Now, we examine the gradient of $\Psi_{AB}$:
\begin{align}
\nabla\Psi_{AB}&=\nabla(\Psi_A\psi_B)=\nabla\Psi_A\cdot\psi_B+\Psi_A\nabla\psi_B\nonumber\\
&=\left(-i\frac{e}{\hbar}\vec{\mathcal{A}}\Psi_A\right)\psi_B+\Psi_A\left(-i\frac{e}{\hbar} \vec{\mathcal{A}}\psi_B\right)\hspace{2.0cm}\rm{from\ Eq.\eqref{eq18}}\nonumber\\
&=-i \frac{2e}{\hbar}\vec{\mathcal{A}}\Psi_{AB}
\end{align}
Similarly, the Laplacian of $\Psi_{AB}$:
\begin{align}\label{eq29}
    \nabla^2\Psi_{AB}&=-i \frac{2e}{\hbar}\nabla\cdot(\vec{\mathcal{A}}\Psi_{AB})\nonumber\\
    &=-i \frac{2e}{\hbar}\vec{\mathcal{A}}\cdot\nabla\Psi_{AB}\nonumber\\
    &=\left(-i \frac{2e}{\hbar}\vec{\mathcal{A}}\right)^2\Psi_{AB}
\end{align}
Therefore, from Eq.\eqref{eq27}, we find that
\begin{align}\label{eq30}
\Psi_{AB}\left(-\frac{\hbar^2}{2m}\nabla^2\psi_q\right) &= -\frac{\hbar^2}{2m} \nabla^2\psi _q'+\frac{\hbar^2}{2m}\left(-i \frac{2e}{\hbar}\vec{\mathcal{A}} \right)^2\psi'_q+\frac{\hbar^2}{2m}2\left(-i\frac{2e}{\hbar}\vec{\mathcal{A}}\right)\Psi_{AB}\cdot(i\vec{k})\psi_q\nonumber\\
&=\frac{1}{2m}\left[-\hbar^2\nabla^2\psi'_q-4e^2\vec{\mathcal{A}}^2\psi'_q+4e\hbar(\vec{\mathcal{A}}\cdot\vec{k})\psi'_q\right]
\end{align}
Like before, we need to rephrase the last term of Eq.\eqref{eq30}. We accomplish this by noticing:
\begin{align}\label{eq31}
    \nabla\psi'_q&=\frac{i}{\bar\eta}\vec{\mathcal{A}}\Psi_A\psi_B\psi_q+\Psi_A\left(-i\frac{e}{\hbar}\vec{\mathcal{A}}\psi_B\right)\psi_q+(i\vec{k}\Psi_A\psi_B\psi_q)\nonumber\\
    &=-2i\frac{e}{\hbar} \vec{\mathcal{A}}\psi'_q + i\vec{k}\psi'_q\nonumber\\
\implies -i\hbar\nabla\psi'_q&=-2e
\vec{\mathcal{A}}\psi'_q+\hbar \vec{k}\psi'_q\nonumber\\
\implies\vec{\mathcal{A}}\cdot\hat{\vec{p}}\psi'_q&=-2e\vec{\mathcal{A}}^2\psi'_q+\hbar(\vec{\mathcal{A}} \cdot\vec{k})\psi'_q\nonumber\\
\implies 4e\hbar(\vec{\mathcal{A}} \cdot\vec{k})\psi'_q&=8e^2\vec{\mathcal{A}}^2\psi'_q+4e\vec{\mathcal{A}}\cdot\hat{\vec{p}}\psi'_q
\end{align}
Using Eq.\eqref{eq31}, the left-hand side of Eq.\eqref{eq30} reduces to:
{\small{
\begin{align}\label{eq32}
    \Psi_{AB}\left(-\frac{\hbar^2}{2m}\nabla^2\psi_q\right) &=\frac{1}{2m}
    \left[-\hbar^2 \nabla^2\psi'_q+4e^2\vec{\mathcal{A}}^2\psi'_q+4e\vec{\mathcal{A}}\cdot\hat{\vec{p}}\psi'_q\right]\nonumber\\
    &=\frac{1}{2m}\left[-\hbar^2\nabla^2\psi'_q
    -4e\cdot i\hbar\vec{\mathcal{A}}\cdot\nabla\psi'_q  +4e^2\vec{\mathcal{A}}^2\psi'_q\right]\nonumber\\
    &=\frac{1}{2m}\left[-\hbar^2\nabla^2\psi'_q
    +\left(-i\hbar\cdot2e\cancelto{0}{(\nabla \cdot\vec{\mathcal{A}})}\psi'_q-i\hbar\cdot2e(\vec{\mathcal{A}}\cdot\nabla)\psi'_q\right)
    -2e\cdot i\hbar\vec{\mathcal{A}}\cdot\nabla\psi'_q  +4e^2\vec{\mathcal{A}}^2\psi'_q\right]\nonumber\\
    &=\frac{1}{2m}(\hat{\vec{p}}+2e\vec{\mathcal{A}})^2\psi'_q
\end{align}
}}
Therefore, we find that the transformed state in a superconductor follows a different Schr$\ddot{\rm o}$dinger equation compared to that in the presence of a magnetic field:
\begin{align}\label{eq33}
    i\hbar\frac{\partial\psi'_q}{\partial t}=\frac{1}{2m}(\hat{\vec{p}}+2e\vec{\mathcal{A}})^2\psi'_q
\end{align}
This is the main justification of why the effective wavefunction can be written as $\Psi_{AB}=\Psi_A\psi_B$. At this point, we also realize that even though we have used a single $q$ in the subscript of $\psi$ in Eq.\eqref{eq25}, actually the charged entity relevant here has a total charge of $-2e$. Note that no property of the charged entity apart from $\psi_q=\int c(\vec{k}){\bf e}^{i(\vec{k}\cdot{\vec{r}}-\omega t)}d\vec{k}$ has been used in the analysis. From this point onwards, we can represent the charged entity in the current model as $\psi_{ee}$. The appearance of $-2e/\hbar$ in the expression of $\Psi_{AB}$ in Eq.\eqref{eq23} may initially appear to be a mathematical coincidence, but Eq.\eqref{eq33} should make it clear that it is not just a coincidence. Also, we must remember that the field wavefunction's description originates in the explanation of the Aharonov-Bohm effect discussed in~\cite{bhattacharya2021demystifying}.
\section{Comparison with existing works}\label{Sec7}
\begin{figure}[H]
    \centering
\includegraphics{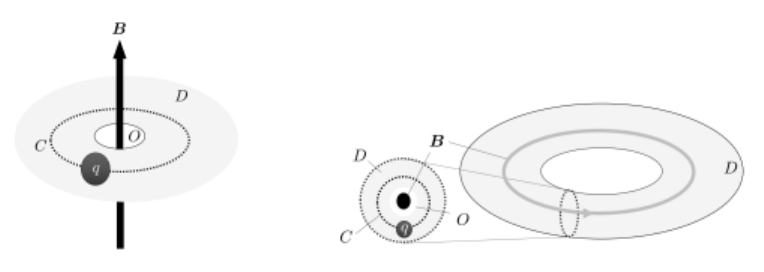}
    \caption{The particle with a charge $q$ moves along the curve $\mathcal{C}$ in a toroidal domain $\mathbf{D}$ where $\vec{\mathcal{B}}={\bf0}$. However, $\vec{\mathcal{B}}\neq{\bf0}$ in the region $\mathbb{O}$. In the right-hand side figure, $\vec{\mathcal{B}}$ is confined in the domain $\mathbb{O}$ by the Meissner effect when $\mathbb{D}$ is a superconductor.}
    \label{f1}
\end{figure}

At this point, it is meaningful to compare this model with some well-established practices in this field. First, we compare the model with the idea proposed by Bayers and Yang~\cite{byers1961theoretical}. Consider a particle of charge $q$ in a domain $\mathbb{D}$ where $ \nabla\times\vec{\mathcal{A}}={\bf 0}$ as shown in figure~\ref{f1}. We have a function $\chi({\bf r})$ on every simply-connected domain such that $\vec{\mathcal A}=\nabla \chi$ by Poincaré’s lemma. We make a multivalued function $\chi({\bf r})$ by laminating them.
We denote a particle of charge $q$ by the wavefunction $\psi_q({\bf r})$. We also define the transformed wavefunction $\psi'_q$ for any wavefunction $\psi_q({\bf r})$ of the particle by $\psi'_q=Exp\left[i\frac{q}{\hbar} \chi({\bf r})\right]\psi_q$. We consider a closed curve $\mathcal {C}$ inside the domain $\mathbb{D} $ surrounding the domain $\mathbb{O}$ through which the magnetic field $\vec{\mathcal{B}}$ passes. Then Bayers and Yang show that:
\begin{align}\label{eq33a}
\psi'_q\rightarrow Exp\left[i\frac{q}{\hbar}\oint_{\mathcal{C}}\vec{\mathcal{A}}\cdot d\vec{s}\right]\psi'_q
\end{align}
Here $\rightarrow$ means that the charged particle rotates once along the curve $\mathcal{C}$. Although we cannot determine the value of the charge $q$ in the above argument, Byers and Yang derive its value and obtain $q=-2$ in the macroscopic arguments using the partition function. They also conclude that the quantization of flux is an indication of the electron pair in the superconductor. The value $q=-2e$ has been confirmed by the experimental result under the conditions in which the Byers-Yang theorems hold as in the right picture of Fig.~\ref{f1}. The experimental result in Refs~\cite{tonomura1986evidence, osakabe1986experimental} not only confirms the Aharonov-Bohm effect but also shows the unit of the flux quantum. Therefore, we realize that the charged particle for the vector
potential $\vec{\mathcal{A}}$ satisfies $q=-2e$:
\begin{align}\label{eq33b}
\psi'_{-2e}\rightarrow Exp\left[-i\frac{2e}{\hbar}\oint_{\mathcal{C}}\vec {\mathcal{A}}\cdot d\vec{s}\right] \psi'_{-2e}
\end{align}
Invoking the condition of single-valued-ness in Eq.\eqref{eq33a} and Eq.\eqref{eq33b}, we can show that the flux is quantized in the units of $\frac{h}{2e}$ within a superconductor. 

To compare this idea to the model proposed in sections~\ref{Sec5} and~\ref{Sec6}, we have to interpret the rotation of the charged particle around the curve $\mathcal C$ once is equivalent to the operation of the wavefunction $\Psi_{AB}$ on the electron wavefunction for a given value of the vector potential $\vec{\mathcal A}$. Since $\mathcal{A}\propto
{\frac{1}{r}}$, we perhaps can say that Eq.\eqref{eq25} is a statement that Eq.\eqref{eq33b} is true for all legitimate values of $\mathcal{A}$ over an set of curves $\mathcal{C}_1, \mathcal{C}_2, ...\mathcal{C}_n$ which can be made into a band as $\mathcal A$ varies continuously. 

In this context, we mention that Malozovsky and Fan~\cite{malozovsky1999magnetic} claim that the factor 2 in $2e$ appearing in the unit of magnetic flux quantum should be the outcome of the London equation and gauge invariance and that it should have nothing to do with the electron pair. They propose that the Aharonov-Bohm effect exists due to the gauge invariance for a particle with a charge $-e$ in an electromagnetic field. In the present model, the gauge invariance is interpreted as a unitary phase acting on the charge states in the absence of fields. However, the present author thinks that the factor of 2 indeed comes from the pairing of electrons in superconductors. This is because we arrived at Eq.\eqref{eq33} with the particle having charge $-2e$, even though we started by assuming a charge $q$. It may be mentioned here that it had been noted at the bottom of page 5 in~\cite{bhattacharya2021demystifying} that for a charge $q$ ($|q|>e$), the phase is given by $Exp\left[i\frac {q}{\hbar}\oint{\vec{\mathcal{A}}\cdot d\vec{s}}\right]$. Secondly, by Noether’s theorem, the global gauge invariance implies the charge conservation (i.e., $\frac{dq}{dt} =0$). Under this gauge invariance, we cannot move from the $(q=-e)$ theory to the $q=-2e$ theory.

Next, we consider certain points by Feynmann~\cite{feynman1965feynman}. He starts from Eq.\eqref{eq11} and considers the continuity equation:
\begin{equation}\label{eq34}
\frac{\partial P}{\partial t} =-\nabla\cdot\vec{\mathcal{J}}
\end{equation}
From Eq.\eref{eq11} and Eq.\eref{eq34}, he finds the current density as:
\begin{equation}\label{eq35}
\vec{\mathcal{J}}=\frac{1}{2}\left[\psi^*\left(\frac{\hat{\mathcal{P}}-q\vec{\mathcal{A}}}{m}\right)\psi+\psi\left(\frac{\hat{\mathcal{P}}-q\vec{\mathcal{A}}}{m}\right)^*\psi^*\right]
\end{equation}
where $m$ is the effective mass of an electron. In the present model, the probability density is defined as $P'=|\psi'_{ee}|^2= \psi'_{ee}{^*}\psi'_{ee}$. This suggests that the time rate of change of the probability density is given by:
\begin{align}\label{eq36}
\frac{\partial}{\partial t}\left|\psi'_{ee}\right|^2
&=\frac{\partial}{\partial t} \left(\psi^*_{ee}\psi^*_B\Psi^*_A\Psi_A\psi_B\psi_{ee}\right)\nonumber\\
&=\frac{\partial}{\partial t} \left(|\Psi_A|^2|\psi_{ee}|^2 \right)
\end{align}
since $\psi^*_B\psi_B=1$. On the other hand, the expression of the current density is still given by Eq.\eqref{eq35}, with a change that $q=-2e$, according to Eq.\eqref{eq33}. The first term of Eq.\eqref{eq35} will be calculated as:
\begin{align*}
{\psi'_{ee}}^*\left(\frac{-i\hbar\nabla+2e\vec{\mathcal{A}}}{m}\right)\psi'_{ee}&={\psi'_{ee}}^*\left(\frac{-i\hbar\nabla\psi'_{ee}}{m}\right)+\frac{2e\vec{\mathcal{A}}}{m}|\Psi_A|^2|\psi_{ee}|^2
\end{align*}
Setting $\psi'_{ee}=\Psi_A\psi_B\psi_{ee}$ and taking clues from Eq.\eqref{eq31}, we find:
\begin{align*}
{\psi'_{ee}}^{*}\left(\frac{-i\hbar\nabla+2e\vec{\mathcal{A}}}{m}\right)\psi'_{ee}&={\psi'_{ee}}^{*}\left(\frac{-2e\vec{\mathcal{A}}\psi'_{ee} -i\hbar\Psi_{AB}\nabla\psi'_{ee} }{m}\right)+\frac{2e\vec{\mathcal{A}}}{m}|\Psi_A|^2|\psi_{ee}|^2\nonumber\\
&=-\frac{2e\vec{\mathcal{A}}}{m}|\Psi_A|^2|\psi_{ee}|^2+\frac {\hbar}{mi}|\Psi_A|^2{\psi_{ee}}^{*}\nabla\psi_{ee}+\frac{2e\vec{\mathcal{A}}}{m}|\Psi_A|^2|\psi_{ee}|^2\nonumber\\
&=\frac{\hbar}{mi}|\Psi_A|^2{\psi_{ee}}^{*}\nabla\psi_{ee}
\end{align*}
Proceeding along similar lines, one can show that the second term in Eq.\eqref{eq35} leads to a term:
\begin{align*}
\psi'_{ee}\left(\frac{-i\hbar\nabla+2e\vec{\mathcal{A}}}{m}\right)^*{\psi'_{ee}}^*&=\psi'_{ee}\left(\frac{i\hbar\nabla{\psi'_{ee}}^*}{m}\right)+\frac{2e\vec{\mathcal{A}}}{m}|\Psi_A|^2|\psi_{ee}|^2\nonumber\\
&=-\frac{\hbar}{mi}|\Psi_A|^2\psi_{ee}\nabla{\psi_{ee}}^*
\end{align*}
From the definition of the current density (Eq.\eqref{eq35}) and the above equations, we get the expression for current as:
\begin{align}\label{eq37}
\vec{\mathcal{J}}=|\Psi_A|^2\cdot\frac{\hbar}{2mi}({\psi_{ee}}^*\nabla\psi_{ee}-\psi_{ee}\nabla{\psi_{ee}}^*)
\end{align}
Which is just $|\Psi_A|^2$ times the common expression for the probability current density of any particle that we find in elementary quantum mechanics. This is simply because we have accounted for the true electronic charge of the state in superconductors. Eq.\eqref{eq36} and Eq.\eqref{eq37} together make the equation of continuity in this context.

Feynmann, however, does not  take $\psi'_{ee}$, rather he represents a pair of electrons as $\psi=\sqrt{\rho(\bf r)}{\bf e}^{i \theta(\bf r)}$, where $\rho$ is the charge density and $\theta$ is the phase of the electron puts $\psi$ in Eq.\eqref{eq35} and shows that 
\begin{equation}\label{eq38}
\vec{\mathcal{J}}=\frac{\hbar}{m}(\nabla\theta-\frac{q}{\hbar}\vec{\mathcal{A}})\rho
\end{equation}
In the present model, if we calculate the current density using $\psi=\psi=\sqrt{\rho(\bf r)}{\bf e}^{i \theta(\bf r)}$, we find that the current density is given by:
\begin{equation}\label{eq39}
\vec{\mathcal{J}}=\frac{\hbar}{m}(\nabla\theta+\frac{2e}{\hbar}\vec{\mathcal{A}})\rho
\end{equation}
Referring to the following figure~\ref{f2}, we assume that the vector potential $\vec{\mathcal{A}}$ leaks out from the toroidal domain $\mathbb{D}$ in the figure and both the electric and the magnetic fields are zero outside $\mathbb{D}$. 
\begin{figure}[H]
    \centering
\includegraphics{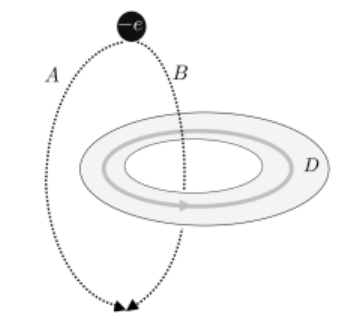}
    \caption{One of the two electron beams from the electron gun passes through the region enclosed by the toroid $\mathbf{D} $ along the path $B$. On the other hand, the other passes outside the toroid $\mathbb{D}$ along the path $A$.}
    \label{f2}
\end{figure}
Since the volume current density (sum of free current $\vec{\mathcal{J}}_{free}$ and the bound current $\vec{\mathcal{J}}_b$) inside a superconductor vanishes, we find that:
\begin{align}\label{eq40}
    \nabla\theta&=-\frac{2e}{\hbar}\vec{\mathcal{A}}\nonumber\\
\implies\oint\nabla\theta \cdot d\vec{s}&=-\frac{2e}{\hbar}\oint\vec{\mathcal{A}}\cdot d\vec{s}
\end{align}
Since we have found that the flux quantum within superconductors are quantized in the units of $\frac{h}{2e}$ [Eq.\eqref{eq24}], the closed loop line integral of the vector potential is the vector potential is nothing but $k\frac{h}{2e}$, where $k$ is an integer. Hence, the above equation becomes:
\begin{align}\label{eq41}
\oint d\theta=\Delta\theta=-2k\pi
\end{align}
This is in contrast with the expression for $\Delta \theta=k\pi$ which one obtains from Eq.\eqref{eq38}. This difference is due to the fact that now we are using two electrons to start with and therefore there is no phase lag between the two electrons represented by the wavefunction. If one chooses to use Eq.\eqref{eq38} (with $q=-e$), it is natural to find $\Delta\theta$ as $k\pi$, because in that case there will be a phase difference of $\pi$ between $|q|=e$ and the electron wavefunction which represents two electrons.
\section{Discussions}\label{Sec8}
The quantized nature of the magnetic flux in the units of $\frac{h}{e}$ has been pointed out by many authors in past~\cite{feynman1965feynman, doll1961experimental, crawford1982elementary}. However, the connection between the usual flux quantum and the flux quantum in superconductivity ($\frac{h}{2e }$) was not very clear on a fundamental ground. The present article uses a simple semi-classical approach to bridge this gap. From this discussion, we find that the result will continue to hold irrespective of type-I, or type-II superconductors, as long as it is in a Meissner state. Elsewhere, it has been argued that the BCS theory cannot explain the Meissner effect~\cite{hirsch2009bcs, hirsch2010explanation}. Actually, the BCS theory can explain the Meissner effect if the $U(1)$ gauge symmetry breaking is taken into account by including the Nambu-Goldstone mode. The references claim that, however, there still exist fundamental problems concerning the way the Meissner effect is realized in realistic situations. Apart from that, it is also difficult to consider the $U(1)$ gauge symmetry breaking since it violates the particle-number conservation. In this context, we comment that if one accepts the flux quantum and possible mechanisms behind it (e.g. the BCS theory), then the present work clearly shows why the magnetic field must be zero inside superconductors.

\subsection{Comments on the Meissner effect}
It is easy to verify that the present model of superconductors is consistent with the Meissner effect. We start with the assertion that the electron wavefunctions in superconductors transform as $\psi_e\rightarrow\psi'_{ee}=\Psi_A\psi_B\psi_{ee} $. Then, we find:
\begin{align}\label{eq42}
    \nabla\psi'_{ee}=\nabla\Psi_A \psi_B\psi_{ee}+\Psi_A\nabla\psi_B\psi_{ee}+\Psi_A\psi_B \nabla\psi_{ee}.
\end{align}
Using Eq.\eqref{eq3}, Eq.\eqref{eq18}, and Eq.\eqref{eq42}, we deduce:
\begin{align}\label{eq43}
\nabla\psi'_{ee}=\frac{i}{\bar\eta} \vec{\mathcal{A}}\Psi_A\psi_B\psi_{ee}+\Psi_A\left(-i\frac{e}{\hbar} \vec{\mathcal{A}}\psi_B\right)\psi_{ee}+&(i\vec{k}\Psi_A\psi_B\psi_{ee})
=-2i\frac{e}{\hbar} \vec{\mathcal{A}}\psi'_{ee} + i\vec{k}\psi'_{ee}\nonumber\\
\implies\nabla(\ln\psi'_{ee})&=-2i\frac{e}{\hbar} \vec{\mathcal{A}}+i\vec{k}
\end{align}
Taking curl on both sides of Eq.\eqref{eq44}, we find:
\begin{align}\label{eq44}
\nabla\times\nabla(\ln\psi'_{ee})&=-2i\frac{e}{\hbar} \nabla\times\vec{\mathcal{A}}+i \nabla\times\vec{k}\nonumber\\
\implies{\bf0}&=\frac{i}{\bar\eta}\vec{\mathcal{B}}+i\cancelto{0}{\nabla\times{\vec{k}}}
\end{align}
Strictly speaking, this is not proof of the Meissner effect in a superconductor. This explanation relies on the magnetic field being zero inside the superconductor. If that were not true, we could not write the eigenvalue equation for $\Psi_A$ or construct $\psi_B$. But this is not an ad-hoc assumption either. There is a clear prediction of the model that the zero field is an indicator of a nonlocal plane wave type state~\cite{bhattacharya2021demystifying}, which in this case, is $\psi_B$. If we accept the nonlocal plane wave nature of $\psi_B$, we can establish the Meissner effect.

\subsection{Comments on Quantum Hall effect}
This work also shows why the observed magnetic flux quantum should be $\frac{h}{e}$ (and not $\frac{h}{2e}$) in say, the integer quantum Hall effect. In this case, the magnetic field is not zero. So, one can have a semi-classical wavefunction $\psi_B$ (leading to one of the factors), but not $\psi_A$ due to a non-zero magnetic field. In fact, it is possible to deduce the transverse resistivity of the integer quantum Hall effect using this simple model. We saw earlier that $\psi_B$ is given by $\psi_B=Exp\left[-i\frac{\oint \mu\mathcal{B}dt}{\hbar}\right]$. This phase can also be expressed as:
\begin{align}\label{eq45}
    \psi_B=Exp\left[-i\frac{\oint \mu\mathcal{B}dt}{\hbar}\right]
    =Exp\left[-i\frac{\oint\frac{e}{2}(xv_y-yv_x)\mathcal{B} dt}{\hbar}\right]
    =Exp\left[-i\frac{\oint\frac{e}{2}(x dy-y dx)\mathcal{B}}{\hbar}\right]
\end{align}
We can define express $xdy-ydx$ in terms of the effective radius $r_0$ of the circulating electrons and $d\phi$.
\begin{align}\label{eq46}
\psi_B=Exp\left[-i\frac{\oint\frac{e}{2}(r_0^2 d\phi)\mathcal{B}}{\hbar}\right]
=Exp\left[-i\frac{\frac{e}{2}(\mathcal{B}r_0^2\oint d\phi)}{\hbar}\right]
=Exp\left[-i\frac{\frac{e}{2}(\mathcal{B}\cdot 2\pi r_0^2)}{\hbar}\right]
\end{align}
Hence, the wavefunction becomes:
\begin{align}\label{eq47}
    \psi_B=Exp\left[-i\frac{e(\mathcal{B}\cdot \pi r_0^2)}{\hbar}\right]
=Exp\left[-i\frac{\Phi_B}{\left(\frac{\hbar}{e}\right)}\right]
\end{align}
where $\Phi_B$ denotes the magnetic flux. The quantum effects manifest when $\Phi_B\sim\frac{\hbar}{e}$. The induced emf due to an increment in the magnetic flux, say due to a change in the magnetic field, is given by $\sim\left|-\frac{\Delta\Phi_B}{\Delta t}\right|\sim\frac{\hbar}{e\cdot \Delta t}$. An expression of the resistivity in $y$ direction is~\cite{tong2016lectures} $\rho_{xy}=\frac{V_y}{I_x}=\frac{\hbar\cdot\Delta t}{e \cdot \Delta t\Delta Q}\sim\frac{\hbar}{e^2\nu}$, where $\nu$ is an integer called the Chern number and heauristically we can say $\Delta Q\sim \nu e$. This relation can be inverted to find the Hall conductivity expression: 
$\sigma_{xy}=\frac{e^2}{\hbar} \nu$. It is also possible to show that the current $I_x$ is quantized using the standard methods. The integer quantum Hall effect is perhaps a manifestation of combined flux and charge quantisation.
\section{Acknowledgement}
The author sincerely thanks the anonymous reviewer for his reading of the manuscript and for raising many questions that led the author to various articles unknown to him and to think deeply about the problem. The discussion in section 7 is a direct fruit of constructive criticism from his end.
\section{Competing interests statement}
The authors have no conflicts of interest to disclose. The work has been accomplished out of academic interests. No funding has been received to carry out the research work.

\section{Data availability statement}
No new data were generated or analyzed in this work.
\section{References}
\bibliographystyle{unsrt}
\bibliography{ePair4SC}
\end{document}